# Information Metrics (*iMetrics*):
# A Research Specialty with a Socio-Cognitive Identity?

**STAŠA MILOJEVIĆ**
*School of Library and Information Science, Indiana University, Bloomington 47405-1901, United States; smilojev@indiana.edu.*

**LOET LEYDESDORFF**
*Amsterdam School of Communication Research (ASCoR), University of Amsterdam, Kloveniersburgwal 48, 1012 CX Amsterdam, The Netherlands; loet@leydesdorff.net*

 **"Bibliometrics", "scientometrics", "informetrics", and "webometrics" can all be considered as manifestations of a single research area with similar objectives and methods, which we call "information metrics" or *iMetrics*. This study explores the cognitive and social distinctness of *iMetrics* with respect to the general information science (IS), focusing on a core of researchers, shared vocabulary and literature/knowledge base. Our analysis investigates the similarities and differences between four document sets. The document sets are drawn from three core journals for *iMetrics* research (*Scientometrics, Journal of the American Society for Information Science and Technology*, and *Journal of Informetrics*). We split *JASIST* into document sets containing *iMetrics* and general IS articles. The volume of publications in this representation of the specialty has increased rapidly during the last decade. A core of researchers that predominantly focus on *iMetrics* topics can thus be identified. This core group has developed a shared vocabulary as exhibited in high similarity of title words and one that shares a knowledge base. The research front of this field moves faster than the research front of information science in general, bringing it closer to Price's dream.**

## Introduction

Terms such as "bibliometrics", "scientometrics", "informetrics", and "webometrics" have been used to describe quantitative studies of bibliographies (books and libraries), science, information phenomena, and the World Wide Web. Although these terms emerged in different contexts and stemmed from different disciplinary backgrounds, they fairly quickly started being used interchangeably. *Bibliometrics*, for example, has its roots in library and information science. The term "bibliometrics" itself was first introduced by Pritchard (1969) to describe "the application of mathematical and statistical methods to books and other media of communication" (p. 348). A number of authors (Borgman & Furner, 2002; Broadus, 1987; White & McCain, 1989) provided their own definitions, some of them (e.g., White & McCain and Borgman & Furner) linking bibliometrics to studies of science and scholarly communication.

*Scientometrics*, defined as the quantitative studies of science (Elkana, Lederberg, Merton, Thackray, & Zuckerman, 1978) or the "quantitative study of science, communication in science, and science policy" (Hess, 1997, at p. 75) has its roots in the 1950s and 1960s and stems from the work of the historian of science Derek de Solla Price (e.g., Price, 1963, 1965) in parallel to the development of the citation indexes by Eugene Garfield (Garfield, 1955, 1963). The first international journal *Scientometrics* specialized in bibliometrics and quantitative studies of science appeared in 1978. In its early years scientometrics has been considered as the quantitative aspect of science and technology studies (STS) that emerged at the similar time (Spiegel-Rösing & Price, 1977). However, STS's main focus was on qualitative sociology of science and research-policy analysis (Leydesdorff & Van den Besselaar, 1997). Courtial (1994), for example, called scientometrics "an hybrid field made of invisible college and a lot of users" (p. 251). With the further development of the science citation index as a powerful new tool, however, scientometrics became increasingly part of the information sciences (IS) during the 1980s and 1990s, and in a number of recent studies scientometrics was considered as part of IS (Lucio-Arias & Leydesdorff, 2009b; Van den Besselaar, 2001; Van den Besselaar & Heimeriks, 2006).

*Informetrics,* defined by Egghe (2005, at p. 1311) as a research area "comprising all-metrics studies related to information science" came into use as a term in the late 1980s (Egghe & Rousseau, 1988). Informetrics can be considered more general than the other two areas, since it includes studies of "the quantitative aspects of information in any form, not just records or bibliographies, and in any social group, not just scientists" (Tague-Sutcliffe, 1992, at p. 1). The expansion of interest in these topics was witnessed by the establishment of a new journal, *Journal of Informetrics,* in 2007. Finally, *webometrics* can be considered as "the application of informetrics methods to the World Wide Web" (Almind & Ingwersen, 1997, p. 404); it is the most recent branch of the four. In 1997 an electronic journal *Cybermetrics* covering primarily webometric research was founded.

Detailed discussion of the similarities and differences among these research areas would lead us away from the objective of this study, and has been extensively covered by others (Hood & Wilson, 2001; Sengupta, 1992). Our general impression is that while these areas of study had different roots, they have evolved to share many of the objectives and have nowadays many methods in common. De Bellis (2009), for example, stated that they are often "indistinguishable". Other studies (e.g., Glänzel & Schoepflin, 1994; Bar-Ilan, 2008) use these terms interchangeably. Given these fundamental similarities and the common focus on *documents* as units of analysis, we consider the four research areas as different labels representing one area of study and call it "*information metrics*" and abbreviate it as "*iMetrics*"[1].

Given this diversity, both in the origins and foci, it is not surprising that many scholars tried to situate the research area that we call *iMetrics* in relation to other fields or disciplines. Thus, a number of studies (Leydesdorff, 2007a; Leydesdorff & Van den Besselaar, 1997; Van den Besselaar, 2000, 2001; Van den Besselaar & Heimeriks, 2006) found strong links between scientometrics, as exemplified by the journal *Scientometrics* and information science, as exemplified by journals such as *Journal of the American Society for Information Science and Technology (JASIST), Journal of Documentation (JDOC),* and *Information Processing & Management (IPM).* These links often lead to the characterization of iMetrics as an integral part of information science (e.g., Åström, 2002; Van den Besselaar & Heimeriks, 2006), but sometimes also as part of "science studies" (e.g., Leydesdorff, 1989; Moya-Anegón, Herrero-Solana, & Jiménez-Contreras, 2006), or at the "cross-roads between science studies and the information science" (Lucio-Arias & Leydesdorff, 2009b: p. 2492). However, other recent studies have noted that, at least in the *cognitive* sense, *iMetrics* can be considered as separate from IS or from the encompassing category of library and information science (LIS). For example, using hierarchical clustering of terms identified from titles in 16 LIS journals Milojević *et al.* (2011) found a strong *iMetrics* branch (exemplified by the journal *Scientometrics*) that stands alongside (and not within) the IS and the library science branches. This reaffirms the finding of Janssens *et al.* (2006) that the journal *Scientometrics* can be largely separated from other LIS journals based on a "different term profile" (p. 1622). The current study further explores the distinctness of *iMetrics* in the cognitive sense, but adds a very important aspect of the *social* distinctness as well.

---

[1] We will also use the term *iMetrics* to discuss the results of the previous studies. We are aware that these authors could not have used this term, since it is introduced in this paper. However, these previous studies often cover the exact same research area which we propose to call here iMetrics for the purposes of brevity and clarity.

For a field of study to be considered a specialty in the sociological sense (Law, 1976; Mullins, 1972), it is not sufficient that it has a distinct cognitive profile, it also needs to have a *social* identity, that is, its practitioners should represent a community whose internal ties are much stronger than the ties with the outside community, even if they institutionally belong to such "outside" social structures in the administrative sense (e.g., belonging to departments or schools of LIS, computer science, etc). Börner et al. (2012) consider research specialty to be "the largest homogeneous unit of science, in that each specialty has its own set of problems, a core of researchers, shared knowledge, a vocabulary and literature" (p. 21). Using bibliometric techniques a research specialty can be operationalized through a study of "an evolving set of related documents" (Lucio-Arias & Leydesdorff, 2009a). The current study examines both the social and the cognitive identity of *iMetrics* as a research specialty with special emphasis on a core of researchers, shared vocabulary and literature.

A number of studies have focused on the nature of what we call *iMetrics*. And while some (e.g., Glänzel & Schoepflin, 1994) considered it to be a field in crisis plagued by the lack of consensus caused by, among other things, the "loss of integrating personalities," others (e.g., Lucio-Arias & Leydesdorff, 2009b; Wouters & Leydesdorff, 1994) have used empirical data to show that *iMetrics* appears to have social identity. For example, in their bibliometric and social network analysis of the journal *Scientometrics* during its first 25 years (1978-1993) Wouters & Leydesdorff (1994) found a coherent well-integrated group of researchers with a cohesive discourse. Van den Besselaar (2000) in the analysis of aggregated journal to journal references for three journals: *Social Studies of Science*, *Scientometrics*, and *Research Policy* found the clustering around *Scientometrics* to be very heterogeneous and to change from year to year, thus prompting him to disagree with Wouters & Leydesdorff's claim that *iMetrics* seems to have formed a stable field. More recently, in their study of four journals (*JASIST, JDOC, IPM* , and *Scientometrics*) applying the analysis of the specific combination of article title words and references Lucio-Arias and Leydesdorff (2009b) found that all these journals except *JDOC* showed an indication of "the interaction at the specialty level" (p. 2495) which is manifested by the similarity in the "topic space".

Some of the differences in these conclusions regarding the status of what we call *iMetrics* stem from not adequately differentiating between the cognitive and social aspects. Other differences can be attributed to the assumptions of the individual studies. Namely, some researchers used *Scientometrics* either as a seed, or the only journal to examine the nature of *iMetrics*, without tying it to any encompassing discipline a-priori (e.g. Wouters & Leydesdorff ,1994), while others assumed *iMetrics* to be part of STS (e.g., Van den Besselaar, 2000, 2001), or information science (e.g., Lucio-Arias & Leydesdorff, 2009b). Finally, some of the differences stem from using journals as units of analysis, and thus not fully considering that journals may often and to varying degrees cover several research areas of which *iMetrics* is one (Boyack & Klavans, 2011).

The goal of the present study is to investigate social and cognitive distinctness of *iMetrics* with respect to general information science. Since distinctness is a relative property one needs a paragon of natural level of heterogeneity present in a research specialty. Thus our analysis revolves around investigating the similarities and differences between four document (i.e., article) sets – three belonging to *iMetrics* and one consisting of articles from general information science without *iMetrics*. The *iMetrics* document sets are drawn from three journals that publish most *iMetrics* research. We then explore similarities between these three *iMetrics* document sets in order to establish the intrinsic level of heterogeneity of *iMetrics* research and then compare each set with non-*iMetrics* documents to establish if and to what extent *iMetrics* research can be considered distinct from IS. Our approach takes advantage of the fact that a large fraction of original and representative *iMetrics* research is published in a very small number of publishing venues—what we call *core iMetrics* journals—because of the prevailing skewedness in scientometric distributions (e.g., Seglen, 1992). Also, we use a relatively straightforward yet effective method to define *iMetrics* and non-*iMetrics* literature which alleviates the issues present in some previous studies of the nature of *iMetrics*.

## Data and methods

*Concept of core iMetrics journals and document sets*

The research area of *iMetrics* has experienced a rapid growth of publication since 1990s (Hood & Wilson, 2001, Van Noorden, 2010). The field is also characterized by specialized journals (Glänzel & Schoepflin, 1994). These are primarily *Scientometrics (SCI)* and *Journal of Informetrics (JOI)*. *Scientometrics* started publishing in 1978 and was the first journal exclusively devoted to the quantitative studies of science. *Journal of Informetrics* is more recent (2007). Specialized journals not only serve to communicate and archive research, but are also a way to establish disciplinary or research field boundaries, a fact that we will use in this study. Furthermore, a significant number of *iMetrics* papers is published in the *Journal of the American Society for Information Science and Technology* (*JASIST*), a journal that also covers more general IS topics. *JASIST* started publishing in 1950 (originally published under the name *Journal of the American Society for Information Science*).[2]

The key to our method is to identify several venues that publish a large number of iMetrics articles, which will define iMetrics datasets whose coherence we explore by comparing them between each other as well as with respect to a non-iMetrics dataset. We derive these datasets from three *core* journals for iMetrics research: *Scientometrics, Journal of Informetrics,* and *Journal of the American Society for Information Science and Technology*. Core journals are very important for the formation of fields by allowing for "coordination of communication and access to reputation,…, knowledge interchange and creation" (Minguillo, 2010; p. 775). We consider the above three journals to be core for *iMetrics* because they publish *most* of the original *iMetrics* research. SCI, JASIST and JOI account for 3/4 of all *iMetrics* papers published in journals classified as LIS in the Journal Citation Reports 2010 of the Web of Science (WoS).

We recognize the fact that there are other journals that publish *iMetrics* articles, most notably *Research Evaluation, Information Processing and Management, Journal of Information Science,* and *Research Policy*. Furthermore, they all have close cognitive ties with the three *iMetrics* journals that we consider core (Leydesdorff, 2007b). However, the volume of *iMetrics* articles published in each of these journals is significantly lower than that in SCI, JASIST and JOI. Applying the same method of distinguishing iMetrics articles that we apply to *JASIST* (described in the next section), we find that *Research Evaluation* has two times fewer *iMetrics* articles that JOI, the smallest of the core journals.

Our choice of core journals is also supported by findings in other studies. For example, in her review of "informetrics" (what we call *iMetrics*) literature, Bar-Ilan (2008) found that *Scientometrics* and *JASIST* have the largest number of informetrics articles, with *Research Policy* in the distant third place (JOI could not be included at the time of her study). Also, we focus on three core journals instead of some larger number because the method we use is based on pair-wise comparisons, so adding many datasets would make the comparisons unwieldy.

Furthermore, we consider *SCI* and *JOI* to be fully specialized in *iMetrics* and therefore use them to operationally define the cognitive domain of *iMetrics*. Using this definition we then split the articles in JASIST using a two-tiered procedure explained in detail in the next section into those belonging to *iMetrics* and those that do not. *iMetrics* articles from JASIST, all articles from SCI and from JOI define the three *iMetrics* document sets. Non-*iMetrics* articles from JASIST define the fourth document set. In this study we will be comparing *iMetrics* document sets among each other (3 comparisons), and each of them to the non-*iMetrics* document set (another 3 comparisons). Note that many studies use entire journals as units of analysis, rather than sets of articles, which then leads to results that are hard to interpret when journals with different breadth of focus are included.

---

[2] Between 1950 and 1970 the journal *JASIS(T)* was published under the title *American Documentation*.

For our study it is essential that we have several, reasonably large, sets of *iMetrics* articles in order to establish the intrinsic level of heterogeneity of *iMetrics*. What is not needed, nor is compatible with our method, is to identify all possible *iMetrics* articles from journals beyond the core ones. The underlying assumption is that the majority of *iMetrics* topics are present in core journals and that most of the active *iMetrics* researchers publish, at least occasionally, in the core journals of this specialty.

*Data defining the document sets*

We downloaded full records from Thomson Reuters' Web of Science for all the publications in SCI, JASIST and JOI; this resulted in 8,280 records.[3] From this set we kept only research articles, by selecting publications classified as "Article" or "Conference paper", as these two document types carry original research results. There were 6,092 such records. For reasons that will become apparent below, only papers published since 1982 were kept in the analysis. JOI began publishing in 2007, which was taken into account where necessary. There were 2,159 and 189 research articles in SCI and JOI, respectively, and they were *ex ante* defined as two of the three *iMetrics* document sets.

To identify *iMetrics* articles in JASIST (third *iMetrics* document set) we employed the following two-tiered procedure. First, we considered any JASIST article that contained references to either SCI or JOI to be an *iMetrics* article, i.e., articles in SCI and JOI are used as the yardstick for the delineation of JASIST articles. This appears to be a reasonable procedure, but one may wonder if it is reliable in cases when, say, only a single reference is made to either SCI or JOI. We checked this by examining the topics of every tenth JASIST article that referenced SCI or JOI only once. All of these articles were found to be unambiguously *iMetrics* related. This citation-based selection yielded 511 *iMetrics* articles from JASIST.

The above method could delineate only JASIST articles that were published since SCI started publishing in 1978. The earliest JASIST article that contains a reference to SCI is from 1982. Thus it apparently took several years for SCI to become 'visible' among the audience publishing in JASIST. Therefore, we restricted all data analysis to the period since 1982, i.e., the period over which the delineation of JASIST articles was possible using this method.

While this citation method provided a very clean sample of *iMetrics* articles in JASIST, it would have missed articles that did not reference SCI or JOI papers, but might nevertheless be considered *iMetrics*. In order to retrieve missing candidate articles we additionally selected post-1982 JASIST papers that contained one of the following seven frequent *iMetrics*-specific words or two prefixes in the title: "citation," "bibliometric," "scientometric," "indicator," "productivity," "mapping," or "cite", as well as the prefixes: "h-" or "co-". We determined the significance of these words and prefixes by analyzing the most frequently occurring words in titles of articles in SCI and JOI after non-specific words and stop-words were omitted. This selection criterion retrieved 81 additional JASIST articles published since 1982. After manually checking which of these candidate papers indeed belonged to *iMetrics*, we removed 19 that did not[4]. Therefore, the final set of *iMetrics* articles published in JASIST consisted of 573 articles that have been very precisely selected employing our mostly unsupervised method. In order to distinguish this set of articles from the general JASIST, we designated it *JASIST-iM*.

---

[3] The data were downloaded on August 20th, 2011.

For establishing the level of socio-cognitive distinctness of *iMetrics* we need a representative and relatively clean sample of *non-iMetrics* articles. We define this final document set to consist of 2,104 JASIST articles published since 1982 that were *not* selected as belonging to *iMetrics*. These are presumably articles covering other aspects of IS.[5] We designated the non-*iMetrics* set of articles *JASIST-O*, where "O" stands for "other".

While our dataset spans three decades (1982-2011), the trends over that period are presented only to set the stage for the remainder of the analysis. The rest of the analysis will focus on the most recent five years (2007-2011)[6]. In other words, we are interested in the properties of the *current* period, and not the dynamics. Building a static picture nevertheless requires a sufficiently long time window such that all the major actors have had a chance to be represented in a structure, yet the period should not be so long as to be affected by changes. The five year period fulfills these criteria. The period since 2007 coincides with the period during which all three core journals have been publishing, and thus allowing us to study them consistently.

In summary, 2,921 research articles published between 1982 and 2011 were identified as *iMetrics* and included in the analysis (573 in JASIST-iM, 2,159 in SCI, and 189 in JOI). For the analysis of the period 2007-2011 we used 1,221 of these 2,921 *iMetrics* articles: 265 from JASIST-iM, 767 from SCI, and 189 from JOI. In addition there were 569 JASIST-O articles during this period.

*Other processing of data*

In order to study social identity of *iMetrics* we disambiguated author names using last names and first initials. The numbers of authors identified in different document sets (both regardless of the placement in the author list, and only first authors) for the period 2007-2011 are provided in Table 1.

| Document set | Articles | Authors | First authors |
|---|---|---|---|
| JASIST-iM | 265 | 359 | 159 |
| SCI | 767 | 1245 | 530 |
| JOI | 189 | 279 | 124 |
| JASIST-O | 569 | 1054 | 435 |
| *iMetrics* | 1221 | 1589 | 686 |

Table 1. Number of different authors (all and first) in four document sets: JASIST-iM, SCI, JOI, JASIST-O, between 2007-2011. Numbers are also given for three *iMetrics* document sets together.

---

[4] Some examples of the papers selected using title keywords, but not belonging to *iMetrics* are: "The representation of national political freedom on web interface design: the indicators", "Does domain knowledge matter: mapping users' expertise to their information interactions", and "Alleviating search uncertainty through concept associations: automatic indexing, co-occurrence analysis, and parallel computing".

[5] This document set may suffer from some "contamination" from unidentified *iMetrics* articles, i.e. those that neither reference SCI or JOI nor feature the nine *iMetrics*-specific words. We were able to estimate the contamination rate of the non-*iMetrics* set to be 4%. This is a tolerable level which cannot be expected to compromise the results of the study.

[6] Note that the data for the final year (2011) was incomplete at the time of this research and will not be taken into account (i.e., 2011 data will be omitted) in trends involving absolute quantities.

Finally, to analyze the topics of articles based on their titles, the following procedure was carried out. We first removed punctuation from titles, and then used software WordStat to consolidate word variants (plurals, etc.). Next, we used WordStat to identify all phrases that occurred three or more times. Phrases can be up to five words long. We then produced a joint list of frequencies of words or phrases (i.e., terms) with stop words and general words excluded. This procedure is explained in Milojević *et al*. (2011). To determine if some term dominates in one document set over others we asked that its frequency in one document set compared to others was larger than 50%. If a term was equally characteristic in all three document sets its contribution in each document set was 33%.

*Measures of (dis)similarity*

To examine the social and cognitive identity of the *iMetrics* and the degree of its distinctness from the general IS community we rely on various measures of similarity. For authors we use simple fractions of authors who are authors in *iMetrics* or non-*iMetrics* document sets other than the one investigated. For other characteristics (title terms and the sources of references) we use cosine similarity. Cosine similarity is a very effective way of establishing the level of (dis)similarity among complex entities (Ahlgren *et al*., 2003). It basically measures the geometrical separation between the multi-dimensional vectors that each represents some property. The smaller the angle between the vectors (the closer the *cosine* is to 1) the more similar they are. If the two vectors are perpendicular (*cosine* = 0) the attributes have nothing in common.

**Results**

*Publication trends in iMetrics document sets*

Before we address the questions of social and cognitive identity of *iMetrics*, let us first examine the overall publishing trends in the three *iMetrics* core document sets. Previous studies (Glänzel & Schoepflin, 1994; Wouters & Leydesdorff, 1994) found increases in the numbers of publications in both JASIST (not just the *iMetrics* articles) and SCI. We update these trends specifically for *iMetrics* articles and show the changes since 1982 (Figure 1).

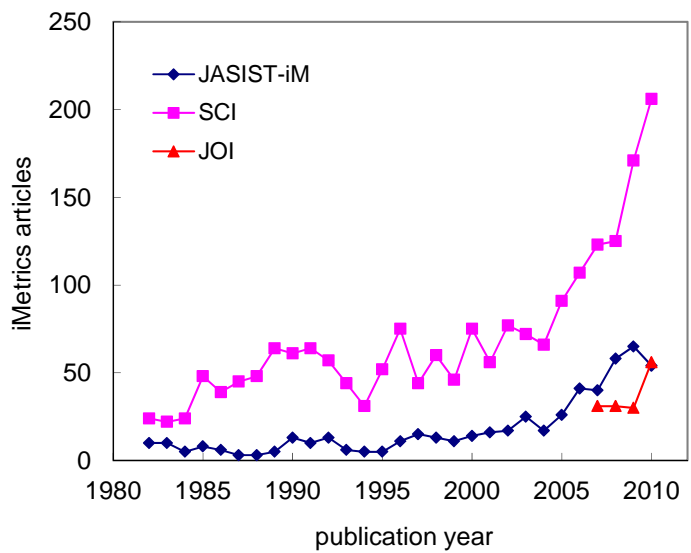

Figure 1. Number of *iMetrics* articles published in JASIST-iM, SCI, and JOI annually since 1982. JOI started publishing only in 2007.

Figure 1 shows an increase in the number of articles in all three *iMetrics* document sets, especially since the mid-90s. In addition, we see that both SCI and JASIST-iM have accelerated the rate of publishing *iMetrics* articles since 2004. Altogether we are witnessing an explosion of the *iMetrics* literature in the core journals, with the number of articles in 2010 approximately four times higher than ten years earlier.

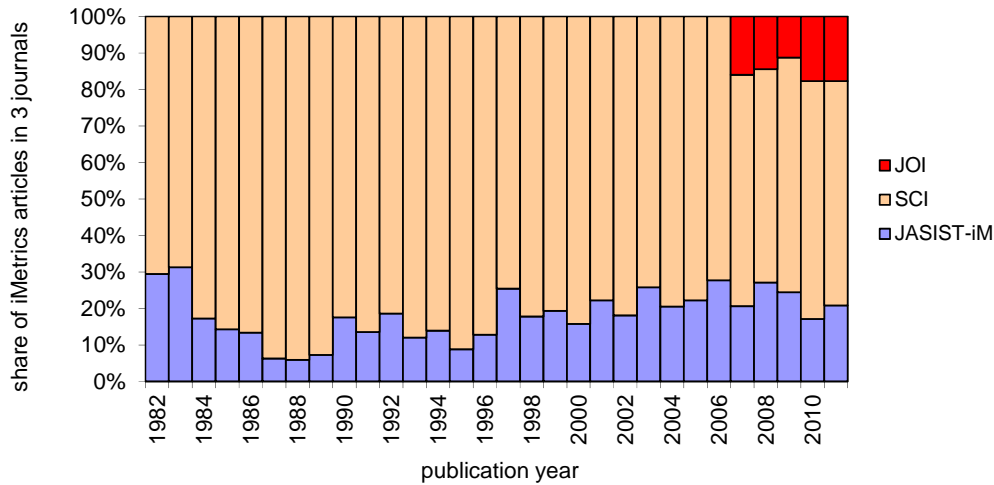

Figure 2. Share of *iMetrics* articles published in JASIST-iM, SCI, and JOI.

Figure 2 compares the shares that each of the three document sets has had in the body of *iMetrics* articles in the three document sets combined. The share of JASIST-iM articles dropped after 1982—perhaps as a consequence of the emergence of SCI—reaching a minimum in 1988. This parallels Lucio-Arias & Leydesdorff (2009b, p. 2494, Figure 6). Since then, however, JASIST has been regaining its share which is now nearly the same as when SCI emerged. JOI appeared in 2007; this led to the decrease in the share of SCI. Currently, SCI publishes approximately half of the *iMetrics* articles in the three document sets, while JASIST-iM and JOI share the other half.

*Social identity and distinctness of iMetrics*

Social identity of *iMetrics* as a research specialty can be studied via the relationships among the researchers. As Crane (1972) and Price (1963) pointed out, science is practiced in fairly close-knit groups of scientists who work on similar problems and who regularly share information with one another. The most visible form of the formal interaction is publication of research articles in journals. In that respect "the interaction of well-defined groups of homogeneous researchers, concentrated around particular sets of journals, leads to the formation of cohesive (sub)groups tied together" (Minguillo, 2010: p. 772). For the reasons already explained, we will not focus our analysis on the relationship between authors and journals, but authors and four different document sets, three of which are considered to be core of *iMetrics* and one is representative of IS. In the analyses we will examine both how strongly interlinked the authors participating in *iMetrics* research are and whether they are distinct from the authors publishing in IS.

To determine if *iMetrics* has social identity, i.e., if the authors publishing *iMetrics* research in core journals are distinct from those who publish on general IS topics, we compare, for each *iMetrics* document set, the fraction of authors who publish in the other *iMetrics* document sets with the fraction of authors who publish outside of *iMetrics*. In Table 2, we use only the data for the most recent five year period (2007-2011).

| Document set | Number of first authors | Fraction of authors who publish in other two iMetrics document sets | Fraction of authors who publish in JASIST-O | Ratio of fractions (coefficient of distinctness) |
|---|---|---|---|---|
| JASIST-iM | 159 | 43% | 14% | 3.1 |
| SCI | 530 | 17% | 4% | 3.9 |
| JOI | 124 | 54% | 7% | 7.4 |

Table 2. Authors in each of the three *iMetrics* document sets and the fraction of them who publish in other two *iMetrics* sets as well as the non-*iMetrics* document set (JASIST-O). Authors are many times more likely to publish in other *iMetrics* venues than in general IS (coefficient of distinctness).

Approximately half of the first authors who publish *iMetrics* articles in JASIST or in JOI also publish (again as first authors) in the other two *iMetrics* venues. So for a large fraction of JASIST-iM and JOI authors those venues are not exclusive.[7] On the other hand, only 17% of SCI first authors also publish in JASIST-iM or in JOI. Such lower percentage is the natural consequence of the fact that SCI is much larger venue for *iMetrics* research than either JASIST or JOI. Consequently, it will be an exclusive venue for a large number of authors. More important in the context of this study is to establish what fraction of authors from the three *iMetrics* document sets also publishes (again as lead authors) in the fourth, non-*iMetrics* document set. Now the fractions are significantly lower (between 4% and 14%).

We can compare the two fractions to determine a *coefficient of distinctness*, i.e., how more likely are the authors to publish in another *iMetrics* venue than in non-*iMetrics* document set (i.e., JASIST-O). This coefficient is 3.1 and 3.9 for JASIST-iM and SCI authors respectively, and as high as 7.4 for JOI authors. The conclusion is that the authors of *iMetrics* research come from the same underlying pool of authors who publish across the board of core *iMetrics* venues. On the other hand, these researchers appear less likely to publish in general IS. This result attests to the high level of social identity and distinctness of *iMetrics* when compared with respect to general IS.

*Cognitive identity and distinctness of iMetrics*

To determine the cognitive identity and the distinctness of *iMetrics* with respect to IS we examine both the article title words and knowledge base as expressed through references.

**Article title words and the cognitive foci of document sets**

First we establish the level of similarity in the cognitive foci of *iMetrics* articles published in the three document sets by analyzing the words that appear in titles of articles (from 2007-2011). The cosine values between the document sets are based on the frequencies of terms that appear in the three sets of titles, but excluding common English words (stop words). The results are provided in Table 3.

---

[7] Since some authors had only had a single publication over the five year time period they will appear as exclusive authors of that document set.

| Document sets | Cosine similarity (error) |
|---|---|
| JASIST-iM & SCI | 0.807 (0.020) |
| JASIST-iM & JOI | 0.830 (0.030) |
| JOI & SCI | 0.779 (0.026) |

Table 3. Cosine similarity between the three *iMetrics* document sets based on the terms used in titles. Values in parentheses are standard deviation errors.

High cosine values (around 0.8) indicate that when it comes to the concepts being used in titles, the cognitive foci of the three *iMetrics* venues are very similar. In terms of the cosine values, JASIST-iM and JOI are somewhat more similar than the two compared to SCI, but the differences are not statistically significant, as can be seen from the errors of the cosine values obtained from bootstrap resampling.

Are such high values of cosine similarity also to be found between the titles of the three *iMetrics* document sets and the titles of general IS articles (JASIST-O)? Results are given in Table 4. The similarity with respect to JASIST-O is much lower (around 0.3) than it is among the *iMetrics* document sets. This confirms that *iMetrics* is distinct with respect to IS in terms of topics present in titles.

| Document sets | Cosine similarity |
|---|---|
| JASIST-iM & JASIST-O | 0.315 |
| JOI & JASIST-O | 0.297 |
| SCI & JASIST-O | 0.301 |

Table 4. Cosine similarity index between the three *iMetrics* document sets and the non-*iMetrics* document set (JASIST-O) based on the terms used in titles.

While there is an overall high level of similarity among the three *iMetrics* venues, we are interested in revealing any specifics in the focus. We approached this problem in two ways. First, we identified the 50 most-frequently used terms in the entire *iMetrics* dataset for the period 2007-2011. We found that the majority (32) of the most-frequently used terms are not dominant in any given document set. Of those that are specific, most belong to SCI because it has the largest share of articles so has the highest contribution to the list of most-frequently used words. Next, we examine 20 most frequent terms that are characteristic for each document set (Table 5). The terms that are overwhelmingly dominant (that is, more than 67% of their occurrence can be attributed to a specific document set) are boldfaced.

| Most frequent characteristic terms from JASIST-iM | Most frequent characteristic terms from SCI | Most frequent characteristic terms from JOI |
|---|---|---|
| citation | **patent** | approach |
| author | performance | evaluation |
| web | collaboration | **type** |
| comparison | china | **application** |
| scholarly | **university** | distribution |
| information | **scientometric** | **core** |
| **access** | international | **empirical** |
| **versus** | country | review |

| open | productivity | g(-index) |
|---|---|---|
| method | **trend** | **hirsch (h-index)** |
| **large** | **nanotechnology** | tool |
| **cocitation** | **authorship** | peer |
| **term** | assessment | output |
| dynamic | **technological** | theory |
| assess | national | **variant** |
| communication | **world** | **bibliographic_couple** |
| subject | evaluate | **word** |
| use | technology | **informetric** |
| result | **r&d** | investigation |
| global | european | level |

Table 5. The list of 20 most specific terms for each document set (2007-2011). The terms that are overwhelmingly dominant are boldfaced.

By taking into account all and not only the overwhelmingly dominant terms we suggest that the specific focus of JASIST-iM can be characterized as the topics related to scholarly communication. In SCI the specific focus is on geographical trends, while in JOI it is on indicators. However, as we have already stated, the similarities between the cognitive coverage of the datasets are much larger than the differences, and the specific terms we have identified point to somewhat higher tendency of appearance of articles on the above topics in those venues, rather than the exclusive coverage of these topics by any of the core venues.

### *Characteristics of the knowledge base in different document sets*

The average number of references per article in SCI and JOI is similar (27 and 30, respectively), but this number is significantly higher in JASIST-iM (40 references). Overall, the *iMetrics* document sets have very similar distributions of the ages of references (Figure 3), with the peak at the age of two years for all three document sets, i.e., most references are recent. Price Indexes[8] are 45, 43 and 51 for JASIST-iM, SCI and JOI respectively. While in JASIST-iM and JOI the number of references that are one year old is almost the same as those at the peak, for SCI this number is much smaller. As a matter of fact, the entire distribution of SCI references appears shifted by approximately one year when compared to those of JASIST and JOI. Rather than attributing this difference to the different practices of authors, an alternative explanation is that it reflects possibly longer times between manuscript submission and publication dates in SCI compared to the other two document sets.

Interestingly, the references in JASIST-O articles are even older than that of SCI. So even though they come from the same journal the Price Index of JASIST-iM articles is 45, while that of JASIST-O articles is 38. In general, we take this to mean that the research front in *iMetrics* moves faster than in the general IS. This is another indication that *iMetrics* has grown cognitively distinct from IS.

---

[8] Price Index (Price, 1970) is the percentage of references (from all articles) up to five years old.

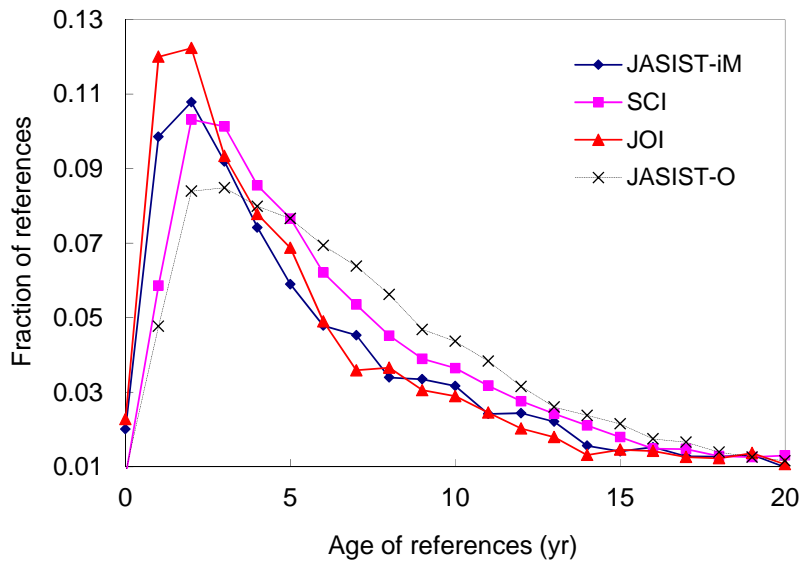

Figure 3. Distribution of the age of references in articles (2007-2011) of a given document set.

Wouters & Leydesdorff (1994) asked themselves whether Price's (1978) dream of "scientometrics" as a "hard" science had come true after 25 years of the journal SCI in existence, and had to answer negatively at the time. Almost two decades later, the situation is unchanged, at least when it comes to the research front of SCI – its Price Index is 43, same as found in Wouters & Leydesdorff (1994). Even JOI, with Price Index of 51, is below the values for "hard sciences" of above 60 (Price, 1970).

Next we explore the make up of the sources that are being referenced in the *iMetrics* document sets (2007-2011). Altogether, there are only nine sources that contribute more than 1% of references in any of the document sets. The list of those sources is presented in Table 6, sorted by the total number of references in all three document sets. The results we obtained are similar to the ones by Peritz and Bar-Ilan (2002) who analyzed the references of articles published in *Scientometrics* (1990-2000) and concluded that "the field relies heavily on itself, on library and information science and on sociology, history and philosophy of science" (p. 282).

| *Source* | *Share in JASIST-iM* | *Share in SCI* | *Share in JOI* |
|---|---|---|---|
| ***JASIS(T)*** | 16.8% | 20.0% | 17.2% |
| ***Scientometrics*** | 12.3% | 18.0% | 14.8% |
| *Res Policy* | 1.5% | *3.9%* | 2.3% |
| ***J Informetr*** | 1.8% | 1.0% | *4.1%* |
| *Science* | 1.3% | 1.5% | 1.7% |
| *P Natl Acad Sci USA* | 1.3% | 1.2% | 2.4% |
| *Nature* | 1.3% | 1.4% | 1.4% |
| *Inform Process Manag* | 1.8% | 0.8% | 1.8% |
| *J Doc* | 1.5% | 0.7% | 0.9% |

Table 6. List of the most frequently cited sources. Percentages in italics are sources predominantly cited by articles in one of the three document sets.

Articles in all three *iMetrics* document sets have been citing JASIS(T)articles (of any type) the most. Interestingly, the papers published in SCI cite papers from JASIST the most of the three. Papers in SCI are most cited by papers in SCI, and papers in JOI by papers in JOI. The contribution of references to JOI in SCI and JASIST is probably lower than its current value because JOI started publishing in 2007, so its full significance cannot be examined yet.

We apply the cosine measure to see how similar/different the knowledge bases of these three document sets are. The results are shown in Table 7. All values are rather high. The largest difference is between JASIST-iM and SCI. JOI, having higher values with respect to both can also be considered as the venue bridging JASIST-iM and SCI.

| Document sets | Cosine similarity |
|---|---|
| JASIST-iM & SCI | 0.879 |
| JASIST-iM & JOI | 0.950 |
| JOI & SCI | 0.947 |

Table 7. Cosine similarity between the three *iMetrics* document sets based on the sources used in references.

Table 8 shows cosines between the three *iMetrics* document sets and JASIST-O. The similarity is again considerably lower than it was among the *iMetrics* document sets. In terms of sources references, JASIST-iM is the most similar to JASIST-O, while SCI is the least similar. It is interesting that JASIST-iM appears to be drawing from the same knowledge base as the general IS, although we have seen that when it comes to the actual topics in titles, it is as dissimilar with respect to JASIST-O as is SCI or JOI.

| Document sets | Cosine similarity |
|---|---|
| JASIST-iM & JASIST-O | 0.589 |
| JOI & JASIST-O | 0.387 |
| SCI & JASIST-O | 0.223 |

Table 8. Cosine similarity index between the three *iMetrics* document sets and the non-*iMetrics* document set (JASIST-O) based on the sources used in references.

Overall, the three *iMetrics* document sets tend to draw from the same knowledge base that is distinct from that of the general IS articles. As an illustration of the similarities, Table 9 lists ten most referenced first author names (during 2007-2011 period) in each of the three *iMetrics* document sets[9]. The lists are similar, with five of the ten names (boldfaced) appearing on all three lists, and two appearing in two venues. All of the authors in this list have also appeared at one time or another as the first authors in *iMetrics* core journals. This indicates that the major contributors to the knowledge base of the field are at the same time active contributors to the core *iMetrics* literature.

---

[9] The cited references in documents downloaded from WoS provide only first author names. The list in Table 9 therefore does not indicate influence or impact of these authors.

| JASIST-iM: | | SCI: | | JOI: | |
|---|---|---|---|---|---|
| **LEYDESDORFF L** | **314** | **GLANZEL W** | **387** | **EGGHE L** | **185** |
| **GARFIELD E** | **210** | **GARFIELD E** | **280** | **GLANZEL W** | **117** |
| **EGGHE L** | **187** | **LEYDESDORFF L** | **259** | BORNMANN L | 111 |
| WHITE HD | 126 | **EGGHE L** | **215** | **LEYDESDORFF L** | **99** |
| **GLANZEL W** | **124** | BRAUN T | 170 | HIRSCH JE | 85 |
| BORNMANN L | 118 | **MOED HF** | **165** | **MOED HF** | **75** |
| CRONIN B | 111 | HIRSCH JE | 140 | **GARFIELD E** | **75** |
| SMALL H | 104 | SCHUBERT A | 132 | ROUSSEAU R | 68 |
| THELWALL M | 91 | NARIN F | 124 | SCHREIBER M | 59 |
| **MOED HF** | **90** | MEYER M | 123 | BURRELL QL | 48 |

Table 9. Comparison of ten most referenced first author names in the three iMetrics document sets (2007-2011). Bold face indicates names appearing in all three sets. Numbers of citations are based on first author names as they appear in reference lists of WoS records and should not be interpreted as evaluative ranking.

**Conclusions**

*iMetrics* is a very active research field experiencing a growth that justifies to talk about an explosion of *iMetrics* literature in the last decade. The number of *iMetrics* articles in core journals in 2010 was some four times higher than ten years before. Whereas during the 1980s and 1990s the *iMetrics* was forming and searching for its identity somewhere between science and technology studies and information science, the research area became more established as it became closer to the information sciences during the 1990s. In the past decade this fusion came to fruition and it is now time to investigate whether *iMetrics* has a full socio-cognitive identity.

While the sheer growth of a research area and the establishment of new venues for publication may suggest the formation of a specialty, for this to actually be the case the practitioners of the research area need to also show signs of social identity, i.e., they need to form a community with ties that are much stronger internally than externally. In order to examine if these criteria are fulfilled in the case of *iMetrics*, we required an appropriate dataset, which we constructed from three core *iMetrics* journals. Assuming that all documents published in SCI and JOI fall into the category of *iMetrics* research, we used these two journals as the yardstick for a two-tiered procedure to identify *iMetrics* papers in JASIST. The division between *iMetrics* and non-*iMetrics* articles in JASIST provided us not only with a clean sample of *iMetrics* documents, but also a comparison group that could be considered as representative of information science research in general.

The approach we used to test for social distinctness of *iMetrics* authors was to compare, for authors of each *iMetrics* venue, the fraction of them who publish in the other two *iMetrics* document sets with respect to the fraction of these authors who publish in non-*iMetrics* document set. We found that most of the *iMetrics* authors are more likely (3 to 7 times) to publish in *iMetrics* document set than in the non-*iMetrics* document set (i.e., JASIST-O). On the other hand, they are typically not tied to any single *iMetrics* venue. These results indicate that the authors of *iMetrics* articles are socially distinct from the more general IS, with only a small fraction working on both the *iMetrics* and non-*iMetrics* topics.

The analysis of topics covered in three *iMetrics* document sets using article title words showed that the differences between the terms are small. On the other hand, the differences with respect to non-*iMetrics* document set are comparatively large. This result points towards very high level of cognitive distinctness. Analyzing the most-frequently used terms that are characteristic for a given document set revealed that in addition to mostly common topics, each venue also has a somewhat specialized focus: for JASIST it is scholarly communication, for SCI studies on particular geographic areas, and for JOI the performance indicators. These differences in foci contribute to some heterogeneity of *iMetrics* research as published in different venues, but the differences are much smaller than the similarities.

The analysis of referencing practices and the knowledge base pointed to further similarities between *iMetrics* venues. Not only do *iMetrics* authors publishing in different venues have similar referencing practices, but the field itself seems to be moving faster than the information science as the reference set. Namely, most references in *iMetrics* papers are recent, they peak at the age of two for all three document sets. The average age of references for non-*iMetrics* papers in JASIST is older.

JASIST is the most referenced source in all three venues. The presence of a shared knowledge base is further supported by the very high values of cosine similarity among the references for the three venues and the comparison of the lists of top ten most cited authors (that shares five authors).

While previous works have strongly indicated that *iMetrics* is a research area with a clearly delineated cognitive focus, we have now shown that *iMetrics* represents a research specialty with a cohesive social and cognitive identity that is distinct with respect to the general information science. The methodology that we have applied or introduced in this work (delineation of multi-topic journals based on citation of single-topic ones, core and comparison document sets) and the associated concepts (notion of the similarity between the core document sets and the distinctness with respect to comparison sets) can also be used to examine if other candidate or proto research areas have achieved social and cognitive identity on the way of developing into full-fledged research specialties or even disciplines.

\*

Acknowledgement
We thank Blaise Cronin and an anonymous referee for providing useful feedback on previous drafts.